\documentclass[conference]{IEEEtran}
\IEEEoverridecommandlockouts

\usepackage{cite}
\usepackage{color}
\usepackage{amsthm,amsmath,amssymb,mathtools,newtxmath}
\usepackage{tabularx}
\usepackage{graphicx}
\usepackage{subfigure}
\usepackage{epstopdf}
\usepackage{amssymb}
\usepackage{amsthm}
\usepackage{algorithm,algpseudocode}

\newtheorem{definition}{Definition}

\begin{document}
\title{Transactive Resilience in Renewable Microgrids: A Contract-Theoretic Approach}
\author{Juntao Chen, Yunhan Huang, and Quanyan Zhu
\thanks{J. Chen is with the Department of Computer and Information Sciences,  Fordham University,  New York,  NY 10023 USA (e-mail: jchen504@fordham.edu).}
\thanks{Y. Huang and Q. Zhu are with the Department of Electrical and Computer Engineering, Tandon School of Engineering,  New York University,  Brooklyn,  NY 11201 USA (e-mail:\ \{yh2315,qz494\}@nyu.edu).}}
\maketitle

\begin{abstract}
Renewable energy-based microgrids play a critical role in future smart grids. Due to the uncertainties of renewable generations, the microgrids face potential risk of load shedding during operation. To address this problem, we propose a contract-based approach to enhance the resilience of microgrids. Specifically, in the framework, the microgrids who may not be self-efficient to meet their local demands can purchase the needed power from their connected microgrids by signing a contract that specifies the power price in advance. We leverage a principal-agent model to capture the energy trading relationships between the microgrids through a resilience as a service (RaaS) paradigm. By focusing on the incentive compatible and individual rational constraints of the service requester, the service provider designs the optimal contracts for the transactive resilience that yields the largest payoff despite the incomplete information. We characterize analytical solutions of the optimal contracts for several scenarios where the service requester has various options on its hidden actions.  Numerical simulations are used to illustrate and corroborate the obtained results.
\end{abstract}


\section{Introduction}
 A microgrid is a green system that relies on renewable distributed resources such as wind turbines, photovoltaics and fuel cells, and it can operate independently from the main power grid in an autonomous manner  \cite{driesen2008design}. A growing number of microgrids are currently interconnected to enhance the system dependability and resiliency \cite{costa2009assessing,colson2011distributed,hussain2019microgrids,chen2017stackelberg}. One benefit of such interconnection is that when one microgrid is out of service, other microgrids in the network can provide it with the necessary power to avoid load shedding and catastrophic cascading failures. Hence, energy transaction between microgrids offers opportunities to maintain the entire power system's stability and resiliency.

One characteristic of a microgrid is that it faces uncertainties from both the renewable generations and the loads, which make its planning and operation a challenging task \cite{mack2014energy}. Based on the predicted load, the grid operators need to decide the amount of power to generate. If the generated power cannot satisfy the real demand, one traditional practice is to cut some loads to balance supply and demand. This operational scheme could lead to dissatisfaction of residents or even blackout in some areas serviced by the corresponding microgrid. A better way to address this issue for the microgrid operator is to purchase a suitable amount of power from other connected microgrids, compensating for self-insufficiency. However, real-time power requests from other agents are  not economical, since the energy price could be tremendously high.

To this end, we propose to develop a contract-based approach \cite{bolton2005contract} that enables direct energy trading between microgrids for system resiliency. Specifically, we offer a resilience as a service (RaaS) framework to capture the energy transactions between interconnected microgrids. In this RaaS framework, there are two types of agents. The first type is the seller or service provider (SP) that provides the resilience service. The other one is the buyer or service requester that requires external power support to avoid load shedding. Under the SaaS framework, the service requester needs to accept the contract specifying the pricing of resiliency service offered by the SP ahead of time before receiving such services. To this end, how to design an appropriate pricing mechanism for transactive resiliency becomes critical.

Both the SP and the resiliency service requester in the established paradigm are strategic. The SP aims to maximize its profits by designing optimal pricing for the resiliency service. The service requester, on the other hand, maximizes its payoff by determining its local generation strategically. One major challenge for designing an efficient contract pricing scheme lies in the asymmetric information between the two involved parties. Specifically, the service requester knows its generation from the traditional means (e.g., generators) and estimates the amount of renewable harvesting. 

In comparison, the SP does not know the real generation of service requester but only the requested power. Designing a RaaS contract by ignoring the strategic behavior of service requester can lead to inefficient operation of the SP. For example, the microgrid will not generate power itself but requests the energy from the RaaS market if the service cost is lower than its own generation cost. To address this challenge, we formulate a mechanism design problem based on a principal-agent model under hidden action for the SP. In the formulation, the SP considers the incentive compatibility and individual rationality of the resiliency service requester explicitly. We design the optimal contracts by using the direct revelation principle \cite{laffont2009theory}.

The main contributions of this paper are summarized as follows.
\begin{enumerate}
\item We propose a contract-based approach to enhance the resiliency of microgrids with renewable generations through a new type of resilience as a service paradigm.
\item We investigate the optimal contract design for transactive resiliency in various scenarios regarding the hidden actions available to the service requester. The designed resiliency contracts also capture the incentives and rationality of the service requester.
\item We use case studies to corroborate the design principles and quantify the benefits of the proposed transactive resiliency approach.
\end{enumerate}

\subsection{Related Works}
Improving critical infrastructure resiliency is a critical research field, and it has been extensively investigated in many sectors, including power systems \cite{li2017networked,chen2016game,zhu2011robust}, transportation systems \cite{leobons2019assessing}, water systems \cite{li2007estimating}, and interdependent infrastructure \cite{cong2018robust,chen2019game}. Unlike the above works, in this paper, we shift our focus from directly improving energy system resiliency through design and control to establishing an energy trading market for transactive resiliency.

Our work is also related to contract mechanism design \cite{bolton2005contract}. Contract-theoretic approaches have been widely used to model the service relationships between the involved parties, with applications to outsourced cloud computing \cite{pham2014optimal}, IoT security \cite{chen2017security}, and demand response in smart grids \cite{haring2016comparing,he2013engage,zhou2018contract}. We leverage a particular type of principal-agent framework under hidden action to design an efficient and incentive compatible pricing scheme for the resiliency services.

\subsection{Organization of the Paper}
The rest of the paper is organized as follows. Section \ref{RaaS} introduces a general RaaS framework for microgrids. Section \ref{PA_problem} formulates the principal-agent problem for RaaS. Two-action and finite-action principal-agent scenarios are analyzed in Sections \ref{two_action} and \ref{finite_action}, respectively. Section \ref{case} presents several case studies, and Section \ref{conclusion} concludes the paper.

\section{Contract-Based RaaS Framework}\label{RaaS}
In this section, we introduce the key components of the contract-based RaaS framework.

\subsection{System Components}
We consider a power system consisting of microgrids that generate renewable energies. The local load of each microgrid $P_l$ is given, and its renewable harvesting $P_r$ is predicted ahead of time and thus faces uncertainty. We only know the probability density function (PDF) $f(P_r)$ of renewable generations which can be determined from history data. In addition to the renewables, microgrid can also generate power from traditional generators, and its amount is denoted by $P_g$ which has no uncertainty. If the total generated power $P_g+P_r$ of microgrid is insufficient to satisfy its local demand $P_l$, then the microgrid needs to buy power from other  microgrids or external main power systems to avoid the load shedding. 

\begin{figure}[!t]
  \centering
    \includegraphics[width=0.55\columnwidth]{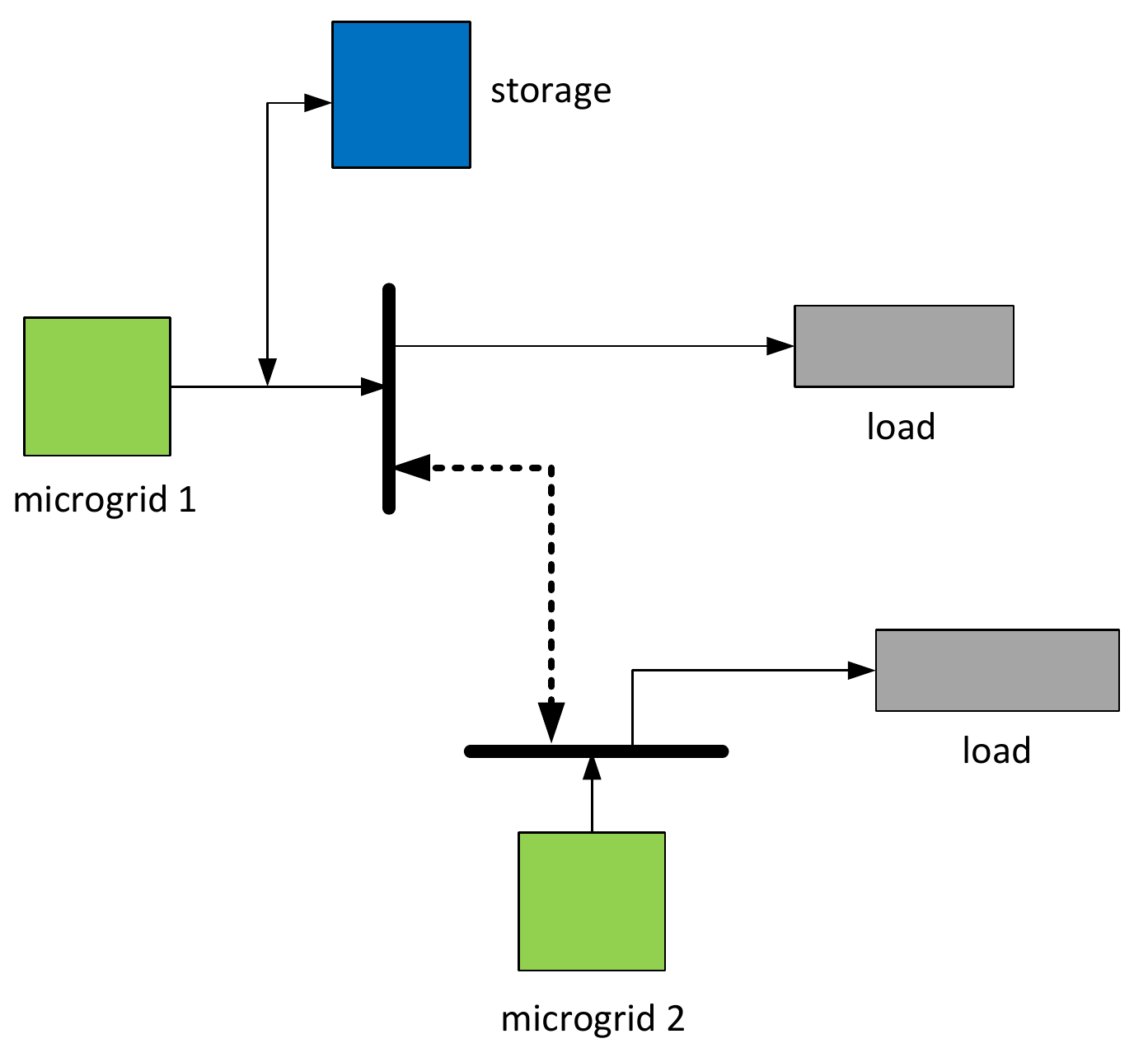}
  \caption[]{RaaS framework in the microgrids. Microgrid 1 will purchase power from microgrid 2 when the local demand is greater than its generations from generators and renewable harvesting. When microgrid 1 generates more power than the actual needed amount, then it stores the remaining to its storage.}\label{model}
\end{figure}

When emergency happens, real-time purchases from utitlity companies to meet local demand are not an economical choice for a mircrogrid. One reason is that the utility company may raise the power price to make more profit due to the emergent large-amount power request from the microgrid. Therefore, a more economical way for the microgrids needs to be designed. In this paper, we use a contract-based approach to capture the resilience service of microgrids. Specifically, the microgrid signs a contract with other agents in the grid on the power price ahead of time. Then, during the real-time operation, the microgrid can buy power from its contractors with a pre-agreed price if its generation does not meet the demand. This resilience as a service paradigm is illustrated in Fig. \ref{model}. The decision variable of the microgrid is the amount of traditional power generation $P_g$. When the microgrid cannot be self-efficient, it requires an amount of $P_l-P_g-P_r$ power support from external grids. In another possible case, the microgrid may generate more power than its demand, i.e., $P_g+P_r>P_l$, and thus it needs to store the excessive $P_g+P_r-P_l$ amount of energy to its storage. 

For convenience, we define $x:=P_l-P_g-P_r$ by the amount of required power from external sources when $x>0$ and the stored power when $x<0$ of the microgrid. Further, denote $P_g^{\max}$ by the maximum traditional generation, and hence the lower bound of $x$ is $\underline{x}=P_l-P_g^{\max}$ and the upper bound is $\bar{x}=P_l$. Specifically, the lower bound $\underline{x}$ corresponds to the scenario that the microgrid prepares for the worst case of renewable harvesting and generates with the maximum power. By default, we assume that the maximum traditional generation is greater than the demand, i.e., $P_l<P_g^{\max}$ and $\underline{x}<0$. The upper bound $\bar{x}$ indicates that the microgrid does not generate traditional power, and the renewable harvesting is also zero. Therefore, the microgrid purely depends on its contractor to satisfy its local load.

\subsection{Incentive Considerations}
The SP's goal is to design a pricing function $H:\mathbb{R}_+\rightarrow\mathbb{R}_+$ such that the microgrid should pay $H(x)$ when requiring $x$ amount of power. We further denote $U(H(x),x)$, $U:\mathbb{R}_+\times \mathbb{R}_+\rightarrow\mathbb{R}_+$, by the net utility of microgrid for receiving the service. Function $U$ includes the benefits for receiving the resilience service and the cost paid for it. The storage cost of microgrid is $S(x)$ which is an increasing function of $x$, where $S:\mathbb{R}_+\rightarrow\mathbb{R}_+$. The traditional power generation cost is denoted by $\Phi(P_g)$, where $\Phi:\mathbb{R}_+\rightarrow\mathbb{R}_+$. Note that $\Phi$ is a continuously increasing and convex function \cite{wood2012power}. In addition, before receiving the service, the microgrid should pay a premium $T$ to the SP for the agreement of delivering services. Note that for given load $P_l$ and generation $P_g$, $x$ shares the same statistics with the renewable harvesting term $P_r$. Denote $f(x|P_g,P_l)$ by the PDF of $x$ given $P_l$ and $P_g$. Therefore, the expected utility of microgrid is 
\begin{equation}\label{Pi_def}
\begin{split}
   \Pi(x,P_g,P_l,H) &:= \int_{0}^{\bar{x}} U\left(H(x),x\right)f(x|P_g,P_l)dx\\
& - \int_{\underline{x}}^0 S(\lvert x \rvert) f(x|P_g,P_l) dx-\Phi(P_g)-T.
\end{split}
\end{equation}

\textit{Remark:} In practice, the density function $f(x|P_g,P_l)$ may not be readily available or not accurate if known. In this scenario, then one may adopt data-driven approach to address this issue. In particular, the first term in \eqref{Pi_def} can be approximated by $\int_{0}^{\bar{x}} U\left(H(x),x\right)f(x|P_g,P_l)dx\approx \frac{1}{N}\sum_{i=1}^N U\left(H(x_i),x_i\right)$, where $x_i$ is data sampling from the distribution $f(x|P_g,P_l)$ and $N$ is the total number of sampling. Here, sampling means the collected data during system oeration. Similarly, we can use $\int_{0}^{\bar{x}} S(|x|)f(x|P_g,P_l)dx\approx \frac{1}{N}\sum_{i=1}^N S(x_i)$ for the second term. In this work, we use a model-based approach and assume that $f(x|P_g,P_l)$ is available.

One natural constraint of the service requester is that it only accepts the designed contract under which it can benefit from the resilience service. This motivates the following definition of individual rationality of microgrid.

\begin{definition}[Individual Rationality]
A contract $H$ satisfies the individual rationality (IR) constraint of the service requester if the following holds:
\begin{equation}\label{IR}
\Pi(x,P_g,P_l,H)\geq 0.
\end{equation}
\end{definition}

The microgrid is strategic in determining its generation level $P_g$ under a contract $H$. Denote $\mathcal{G}$ by a compact set including all possible $P_g$. Then, the rational behavior of microgrid can be captured by the incentive compatibility constraint as follows.

\begin{definition}[Incentive Compatibility]
The action $P_g$ adopted by the micorgrid is incentive compatible (IC) if it maximizes the expected utility $\Pi$, i.e.,
\begin{equation}\label{IC}
P_g\in \arg\max_{P_g'\in\mathcal{G}}\ \Pi(x,P_g',P_l,H).
\end{equation}
\end{definition}

Note that a contract is incentive feasible if it satisfies IR constraint \eqref{IR} and IC constraint \eqref{IC}.

Buying power from external grids based on the contract enhances the resilience of microgrid. The acquired resilience can be quantified by the net benefit from the avoidance of load shedding without the service. The resilience metric adopted by our framework is introduced in Section \ref{resilienct_metric}.

\subsection{Resilience Service Flow}
It is worth discussing the timing of events involved in the proposed RaaS framework.  The resilience contract design can be divided into two stages: the contracting stage and the execution stage. At the first stage, the SP designs an incentive compatible contract that specifies the unit power price as well as a premium fee in entering this transactive resilience market. The service requester chooses whether to accept the offered contract, and this completes the contracting stage. If the service requester agrees with the contracted terms, then the service relationship between these two parties is established. In the execution stage, when there is a need for the resilience service, the service requester can request the SP, and the corresponding service fee (i.e., power price) is provided by the agreed contract.

\section{Principal-Agent Model for RaaS}\label{PA_problem}
In this section, we formulate the RaaS problem based on the principal-agent (PA) model with hidden actions. We further define the resilience metric and characterize the challenges in solving the PA problem.

\subsection{PA Model with Hidden Actions of Microgrid}
In our established RaaS framework, the amount of renewable energy harvesting $P_r$ of microgrid faces uncertainty, and its traditional power generation $P_g$ is a hidden decision variable.  Due to the hidden action of microgrids, we use a principal-agent\footnote{The principal-agent model under hidden action structure is also called moral hazard in contract theory. In this paper, the principal refers to the SP, and the agent refers to the service requester.} model to capture the strategic behaviors of the power requester and SP. Detailed introduction of the PA model can refer to \cite{bolton2005contract,laffont2009theory,
caillaud2000hidden}. Specifically, the microgrid's hidden action $P_g$ belongs to a compact and continuous set $\mathcal{P}_g$. Then, to design an optimal contract $H$, the SP solves the following hidden action problem:
\begin{align*}
\mathrm{(PA):}\ \max_{H(\cdot),P_g}&\quad \int_{0}^{\bar{x}} V\left(H(x),x \right)f(x,P_g,P_l) dx \\
&\mathrm{s.t.}\quad \eqref{IR}\ \mathrm{and}\ \eqref{IC},
\end{align*}
where $V:\mathbb{R}_+\times \mathbb{R}_+\rightarrow\mathbb{R}_+$ is the payoff of the SP which is a function of the received payment $xH(x)$ and service delivery cost.  The limit $x\in[0,\bar{x}]$ in the objective function indicates that the microgrid purchases power from the SP.

The IR constraint \eqref{IR} ensures that the service requester will not accept the contract provided by the SP if his minimum benefit $T$ is not met. In addition, the IC constraint \eqref{IC} indicates that the service requester is strategic by selecting the best action for a given contract.

\subsection{Resilience Metric and Value of Information}\label{resilienct_metric}
Based on the designed contract $H(x)$ with other agents, the microgrid can avoid $x$ amount of load shedding when $x>0$. Thus, the enhanced resilience $R(H(x),x)$ through contract $H(x)$ of microgrid when requiring $x>0$ amount of power support can be quantified by following:
\begin{align}
R(H(x),x) = U(H(x),x)-T,\ x>0.
\end{align} 
Similarly, the average resilience enhancement of using the service by taking into account the renewable generation uncertainty can be quantified by $\bar{R}$ which is defined as follows:
\begin{equation}\label{resilience_def}
\begin{split}
   \bar{R}(x,H,P_g,P_l) = \int_{0}^{\bar{x}} U\left(H(x),x\right)f(x|P_g,P_l)dx-T.
\end{split}
\end{equation}

In the contract mechanism design, the resilience SP anticipates the adopted action of the service requester under a given contract. Therefore, determining the possible actions $P_g$ of the service requester is imperative. Specifically, the definition of implementable action is as follows.
\begin{definition}
An action $P_g$ is implementable if there exists at least one contract $H(\cdot)$ that satisfies the IC and IR constraints.
\end{definition}

When the action of microgrid is observable and verifiable by the SP, then the problem solved by the principal becomes
\begin{align*}
\mathrm{(PA'):}\ \max_{H(\cdot),P_g}&\quad \int_{0}^{\bar{x}} V\left(H(x),x \right)f(x,P_g,P_l) dx \\
&\mathrm{s.t.}\quad \eqref{IR}.
\end{align*}
Note that the IC constraint does not appear in the above problem, since the principal can verify the adopted action by the agent. 
The value of information (VoI) can be defined as the difference of  utilities that the SP can obtain under the optimal contracts resulting from $\mathrm{(PA)}$ and  $\mathrm{(PA')}$, i.e.,
$
\mathrm{VoI} := \int_{0}^{\bar{x}} \big[V\left(H^o(x),x \right)f(x,P_g^o,P_l)
- V\left(H^*(x),x \right)f(x,P_g^*,P_l)\big] dx,
$
where $(H^o(\cdot), P^o_g)$ and  $(H^*(\cdot), P^*_g)$ are the solutions to problems $\mathrm{(PA')}$ and  $\mathrm{(PA)}$, respectively.

\textit{Remark:} Without the IC constraint, the optimal objective of $\mathrm{(PA')}$ is no less than the one resulting from $\mathrm{(PA)}$. Therefore, it is straightforward to obtain $\mathrm{VoI}\geq 0$.

\subsection{Challenges of Designing Optimal RaaS Contracts}
Assume that the first order approach is valid. Then, the first order condition (FOC) of IC constraint \eqref{IC} yields
\begin{equation}\label{FOC}
\begin{split}
&\int_{0}^{\bar{x}} U(H(x),x)f_{P_g}(x,P_g,P_l) dx \\
&+ \int_{\underline{x}}^0 S(|x|) f_{P_g}(x,P_g,P_l)dx -  \Phi'(P_g) = 0.
\end{split}
\end{equation}
Thus, we can replace the IC constraint \eqref{IC} with \eqref{FOC}. The Lagrangian of (PA) is 
\begin{equation}\label{Lagrangian}
\begin{split}
\mathcal{L}(H,P_g,\lambda,\mu)&=\int_{0}^{\bar{x}} V(H(x),x)f(x,P_g,P_l) dx \\
&+ \lambda\Big( \int_{0}^{\bar{x}} U(H(x),x))f(x,P_g,P_l)dx \\
&- \int_{\underline{x}}^0 S(\lvert x \rvert) f(x,P_g,P_l) dx - \Phi(P_g) - T \Big)\\
&+\mu\Big( \int_{0}^{\bar{x}} U(H(x),x)f_{P_g}(x,P_g,P_l) dx \\
&+ \int_{\underline{x}}^0 S(|x|) f_{P_g}(x,P_g,P_l)dx -  \Phi'(P_g) \Big),
\end{split}
\end{equation}
where $\mu$ and $\lambda$ are nonnegative Lagrange multipliers of IR \eqref{IR} and IC \eqref{IC} constraints, respectively.
By using FOC to \eqref{Lagrangian} with respect to $H(x)$, we obtain
$
\frac{\partial V}{\partial H}f(x,P_g,P_l) + \lambda \frac{\partial U}{\partial H} f(x,P_g,P_l)
-\mu \frac{\partial U}{\partial H} f_{P_g}(x,P_g,P_l)=0,
$ $\forall x$.
However, \eqref{FOC} usually does not hold for the normal distribution of $f(x,P_g)$ since the decision-makings $H(\cdot)$ and $P_g$ in (PA) are \textit{coupled}, and they should be addressed simultaneously. Therefore, in Sections \ref{two_action} and \ref{finite_action}, we investigate several special scenarios of the PA problem to obtain more insights of  RaaS for microgrids.

\section{Two Actions Scenario}\label{two_action}
In this section, we focus on the cases that the agent can take two actions including $P_g^L$ and $P_g^H$, $P_g^H>P_g^L$, which correspond to low-level and high-level generations, respectively. The practical interpretations of these two actions are as follows. The microgrid takes action $P_g^L$ based on the maximum expected renewable generation, while takes action $P_g^L$ according to the mean value of expected renewable harvesting. Furthermore, these two actions may induce a finite number of outcomes\footnote{In this paper, the term \textit{outcome} refers to the amount of requested power of the microgrid.}, i.e.,  power request, depending on the service provided by SP. Specifically, in Section \ref{two_action_two_outcome}, we investigate the scenario that the SP provides two types of resilience services. We further consider more options of resilience services in Section \ref{two_action_multi_outcome}.

\subsection{Two Actions and Two Outcomes Case}\label{two_action_two_outcome}
Knowing that the microgrid can only generate $P_g^L$ or $P_g^H$ level of energy, and we denote the corresponding outcomes by $x_H$ and $x_L$,  respectively. In particular, we obtain $x_H>x_L>0$ such that the microgrid is not self-efficient and requires power from the resilience SP. Thus, the contract includes two terms $H_{x_H} := H(x_H)$ and $H_{x_L} := H(x_L)$. In addition, the uncertainty of renewable energy harvesting is captured by the distribution $F(x_L,a) = k-q\mathbf{I}_{\{a=P_g^L\}}$, where $\mathbf{I}_{\{\cdot\}}$ is an indicator function; $q\in(0,1]$ and $k\in[q,1]$ are known constants. $F(x_L,a)$ yields the probability that the microgrid requires $x_L$ amount of power from SP when taking action $a$. Then, the contract design problem can be captured by the following two problems:
\begin{align*}
\mathrm{(PA-1}&\rm{-a)}:\\
 \max_{H_{x_H},H_{x_L}}&\ (1-k)V(H_{x_H},x_H) +kV(H_{x_L},x_L)\\
\mathrm{s.t.}\ &(1-k)U(H_{x_H},x_H) +kU(H_{x_L},x_L)-\Phi(P_g^H) \geq T,\\
& (1-k)U(H_{x_H},x_H) +kU(H_{x_L},x_L)-\Phi(P_g^H)\geq \\
(1+q&-k)U(H_{x_H},x_H) + (k-q)U(H_{x_L},x_L)-\Phi(P_g^L).\\
\mathrm{(PA-1}&\rm{-b)}:\\
 \max_{H_{x_H},H_{x_L}}&\ (1+q-k)V(H_{x_H},x_H) + (k-q)V(H_{x_L},x_L)\\
\mathrm{s.t.}\ &(1+q-k)U(H_{x_H},x_H) +(k-q)U(H_{x_L},x_L)\\
&\qquad\qquad\qquad\qquad\qquad\qquad\ -\Phi(P_g^L) \geq T,\\
(1+q&-k)U(H_{x_H},x_H) + (k-q)U(H_{x_L},x_L)-\Phi(P_g^L)\\
&\geq (1-k)U(H_{x_H},x_H) +kU(H_{x_L},x_L)-\Phi(P_g^H).
\end{align*}
Specifically, (PA-1-a) and (PA-1-b) correspond to the cases that the resilience SP anticipates the service requester generating $P_g^H$ and $P_g^L$ amount of power, respectively. By comparing the optimal contracts of (PA-1-a) and (PA-1-b), the SP announces the one that yields higher payoff. 

To facilitate the optimal contract design, we adopt $V\left(H(x),x \right)=H(x)-\zeta x$, $\Phi(P_g)=\alpha P_g^2+\beta P_g+\gamma$, where $\zeta>0$ is the unit cost of delivering service, and $\alpha,\ \beta,\ \gamma$ are positive constant parameters. In addition, the utility function is $U\left(H(x),x \right)=\psi(x)- {H(x)}$, where $\psi:\mathbb{R}_+\rightarrow \mathbb{R}_+$ captures the satisfaction of microgrid serving its load successfully. Note that $\psi$ is considered to be monotonically increasing. 
 Then, (PA-1-a) is equivalent to
\begin{align*}
\max_{H_{x_H},H_{x_L}}&\quad (1-k) H_{x_H}+k H_{x_L}\\
\mathrm{s.t.}&\quad  H_{x_H} - H_{x_L}  \geq \psi(x_H)-\psi(x_L)+\frac{\Phi(P_g^H)-\Phi(P_g^L)}{q},\\
&\quad (1-k) H_{x_H}+k H_{x_L} \leq (1-k)\psi(x_H)\\
&\qquad\qquad\qquad\qquad+k\psi(x_L)-\Phi(P_g^H)-T.
\end{align*}
One optimal solution to the above linear program can be obtained by solving two inequality constraints with equality jointly (both the IC and IR constraints are binding), i.e.,  
\begin{align}
& H_{x_H} -  H_{x_L} = \frac{\Phi(P_g^H)-\Phi(P_g^L)}{q}-\psi(x_L)+\psi(x_H),\notag\\
& (1-k)H_{x_H}+k  H_{x_L} = (1-k)\psi(x_H)+k\psi(x_L)-\Phi(P_g^H)-T,\notag\\
&\Rightarrow \begin{cases}
H_{x_H} = \psi(x_H)-\Phi(P_g^H)+k \frac{\Phi(P_g^H)-\Phi(P_g^L)}{q}-T,\\
 H_{x_L} = \psi(x_L)-\Phi(P_g^H)-(1-k)\frac{\Phi(P_g^H)-\Phi(P_g^L)}{q}-T.
\end{cases}\label{two_action_solution}
\end{align}
We can verity that \eqref{two_action_solution} is an optimal solution to (PA-1-a). Since $ \psi(x_H)-\psi(x_L)+\frac{\Phi(P_g^H)-\Phi(P_g^L)}{q}>0$, the service price $H_{x_H}$ is always greater than $H_{x_L}$.  In addition, the SP's profit is $(1-k)(\psi(x_H)-\zeta x_H)+k(\psi(x_L)-\zeta x_L)-\Phi(P_g^H)$.

Similarly, we can obtain the solution to (PA-1-b) as 
\begin{align}
\begin{cases}
H_{x_H} = \psi(x_H)-\Phi(P_g^H)+k \frac{\Phi(P_g^H)-\Phi(P_g^L)}{q}-2T,\\
 H_{x_L} = \psi(x_L)-\Phi(P_g^H)-(1-k)\frac{\Phi(P_g^H)-\Phi(P_g^L)}{q}-2T.
\end{cases}\label{two_action_solution_2}
\end{align}
Note that the service price $H_{x_H}$ is also greater than $H_{x_L}$ in this optimal contract, and the SP's payoff is $(1+q-k)(\psi(x_H)-\zeta x_H)+(k-q)(\psi(x_L)-\zeta x_L)-\Phi(P_g^L)-T$. Therefore, by comparing the optimal payoffs in (PA-1-a) and (PA-1-b), we obtain the following result. When 
$$q(\psi(x_L)-\psi(x_H)-\zeta (x_L-x_H))+\Phi(P_g^H)-\Phi(P_g^L)+T>0,$$
then the SP implements the contract in \eqref{two_action_solution}, and otherwise the one in \eqref{two_action_solution_2}.
Sensitivity of the parameter $q$ on the contract will be investigated numerically in Section \ref{case}.

\textit{Remark:} Note that $\mathrm{VoI}=0$ in this two-action two-outcome scenario, since the IR constraint in (PA-1-a) and (PA-1-b) determines the optimal payoff of the SP, and it also appears in the full-information problem as shown in $(\mathrm{PA}')$.

\subsection{Two Actions and Multiple Outcomes Case}\label{two_action_multi_outcome}
We consider a more general multiple-outcome case, i.e., $x$ is in a finite set $\mathcal{X}$. The storage cost is captured by $S(x)=\tau x$, where $\tau>0$ denotes the unit cost. When the SP prefers the microgrid to generate $P_g^H$ amount of power, the PA problem becomes 
\begin{align}
&\mathrm{(PA-2)}:\notag\\
\max_{H(\cdot)}&\quad \sum_i f(x_i,P_g^H,P_l) \left(H(x_i)-\zeta x_i \right)\notag\\
\mathrm{s.t.}&\quad \sum_{i\in\{i:x_i>0\}} \left(\psi(x_i)-{H(x_i)}\right)f(x_i,P_g^H,P_l) \label{two_continuous_IR}\\
&- \sum_{i\in\{i:x_i\leq 0\}} \tau\lvert x_i \rvert f(x_i,P_g^H,P_l) 
- \Phi(P_g^H)\geq T,\notag \\ 
&\sum_{i\in\{i:x_i>0\}} \left(\psi(x_i)-{H(x_i)}\right)f(x_i,P_g^H,P_l) \notag\\
&- \sum_{i\in\{i:x_i\leq 0\}} \tau\lvert x_i \rvert f(x_i,P_g^H,P_l) 
- \Phi(P_g^H)\notag \\
&\geq \sum_{i\in\{i:x_i>0\}} \left(\psi(x_i)-{H(x_i)}\right)f(x_i,P_g^L,P_l) \label{two_continuous_IC}\\
&- \sum_{i\in\{i:x_i\leq 0\}} \tau\lvert x_i \rvert f(x_i,P_g^L,P_l) 
- \Phi(P_g^L)\notag .
\end{align}
Note that (PA-2) is a linear program with decision variables $H(x_i)$, where $x_i>0$ and $x_i\in\mathcal{X}$. The IR constraint \eqref{two_continuous_IR} is binding at the optimal solution since otherwise, the principal can increase the service price to earn more profits while still making the constraint satisfied. We can also formulate the PA problem when the SP anticipates the microgrid generating $P_g^L$ amount of power in a similar way. As the two-outcome case in Section \ref{two_action_two_outcome}, the final optimal contract is the one that yields a higher payoff of the SP.

Next, we investigate the impact of storage associated with the agent on the contract design. By focusing on the binding IR constraint in (PA-2), i.e.,
\begin{align*}
&\sum_{i\in\{i:x_i>0\}} \left(\psi(x)-{H(x)}\right)f(x,P_g^H,P_l)dx \\
&\quad =  \sum_{i\in\{i:x_i\leq 0\}} \tau\lvert x \rvert f(x,P_g^H,P_l) dx + \Phi(P_g^H) + T,
\end{align*}
we obtain that when the unit storage cost $\tau$ is larger, the unit power price $H(x)$ in the contract should be lower. Otherwise, the IR constraint cannot be met, and the agent does not accept the designed contract. 

\section{Finite Actions Scenario}\label{finite_action}
In this section, we generalize the two-action case to a scenario that the microgrid has a finite number of choices of local generation $P_g\in\mathcal{P}_g$, and the corresponding outcome $x$ is in a finite set $\mathcal{X}$. This modeling indicates that the microgrid has a larger freedom in its local generation and the contract offers various options of resilience services.

We denote the outcome by $x_j$, where $j\in\mathcal{J}:=\{1,...,J\}$, and without loss of generality, we have $x_i\leq x_j$ if $i<j$. Under the countable number of possible outcomes, the principal designs the resilience contract by solving the following problem, for each $P_g^\dagger\in\mathcal{P}_g$,
\begin{align*}
&\mathrm{(PA-3)}:\\
\max_{H(\cdot)}&\quad \sum_{i\in\{i:x_i>0\}} f(x_i,P_g^\dagger,P_l)H(x_i) \notag\\
\mathrm{s.t.}&\  \sum_{i\in\{i:x_i>0\}} \left(\psi(x_i)-{H(x_i)}\right)f(x_i,P_g^\dagger,P_l) \\
&- \sum_{i\in\{i:x_i\leq 0\}} \tau\lvert x_i \rvert f(x_i,P_g^\dagger,P_l) 
- \left( \alpha {(P_g^\dagger)}^2+\beta P_g^\dagger+\gamma \right)\geq T,\\
&\sum_{i\in\{i:x_i>0\}} \left(\psi(x_i)-{H(x_i)}\right)f(x_i,P_g^\dagger,P_l) \\
&- \sum_{i\in\{i:x_i\leq 0\}} \tau\lvert x_i \rvert f(x_i,P_g^\dagger,P_l) - \left( \alpha {(P_g^\dagger)}^2+\beta P_g^\dagger+\gamma \right)\\
&\geq \sum_{i\in\{i:x_i>0\}} \left(\psi(x_i)-{H(x_i)}\right)f(x_i,P_g,P_l) \\
&- \sum_{i\in\{i:x_i\leq 0\}} \tau\lvert x_i \rvert f(x_i,P_g,P_l) - \left( \alpha {(P_g)}^2+\beta P_g+\gamma \right),\\
&\qquad\qquad\qquad\qquad\qquad\qquad\qquad\qquad\quad \forall P_g\in\mathcal{P}_g.
\end{align*}
Note that (PA-3) is a convex optimization problem which can be solved efficiently. It is also worth mentioning that the SP should address $|\mathcal{P}_g|$ problems to determine the optimal resiliency pricing.
For convenience, we denote the implementable set of actions as $\mathcal{P}_g^I\subseteq \mathcal{P}_g$. We next develop a two-step approach to characterize the optimal solution in this scenario. The main idea is as follows. First, the SP designs the optimal contract that implements a given action of the microgrid. Then, the SP determines the  action to implement which yields the best payoff.  
For clarity, the two-step approach is summarized as follows.
\begin{enumerate}
\item For each $P_g\in\mathcal{P}_g^I$, the principal solves (PA-3) and denote $B(P_g)= \max_{H(\cdot)}\ \sum_{i\in\{i:x_i>0\}} f(x_i,P_g,P_l)H(x_i)$. In addition, denote $C(P_g) = \sum_{i\in\{i:x_i>0\}} f(x_i,P_g,P_l)\zeta x_i$ as the cost that principal provides the service when the agent adopts action $P_g$.
\item The principal chooses a contract such that the agent implements $P_g^* \in \arg\max_{P_g}\ B(P_g) - C(P_g)$.
\end{enumerate}
When the number of outcomes $J$ increases, then the SP needs to specify more terms in the contract on the service prices. Therefore, we can approximate the continuous outcome case of microgrid with a sufficiently large $J$.

\section{Case Studies}\label{case}
In this section, we use several case studies to illustrate the obtained results. The benefit of resilience service is captured by the function $\psi(x)$ which is monotonically increasing over $x$. For illustration purpose, we choose $\psi(x)=\kappa x^{\rho}$, where $\kappa$ and $\rho$ are positive weighting constants.

\subsection{Case Study 1: Two-Action Two-Outcome}
\begin{figure}[!t]
\centering
\includegraphics[width=0.8\columnwidth]{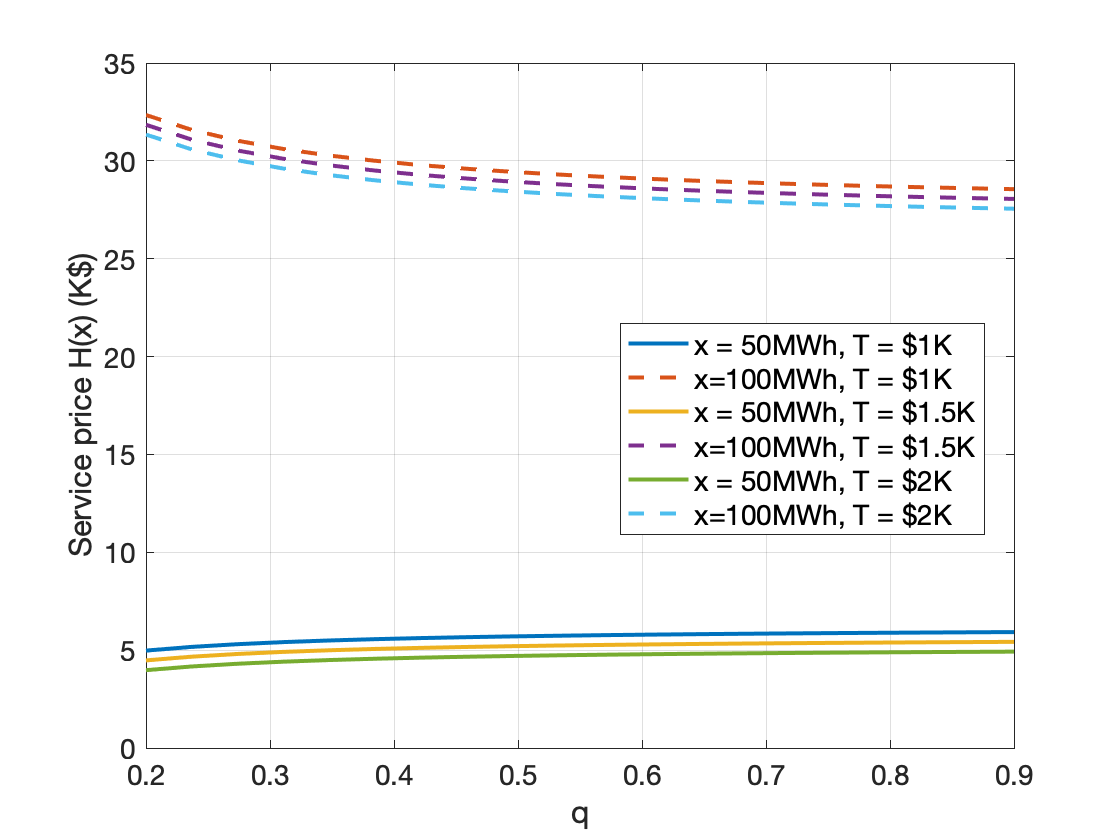}
\caption{Resilience service fee for receiving $x_L$ and $x_H$ amount of energy from the SP with respect to different parameter $q$.}\label{two_action_price}
\end{figure}

We first consider two actions and two outcomes in this case study. The parameters are as follows: $P_g^L=200\rm{MWh}$, $P_g^H=240\rm{MWh}$, $x_L = 50\rm{MWh}$, $x_H = 100\rm{MWh}$, $k=0.8$, and other parameters are summarized in Table \ref{t1}. We study the optimal contract design under the regime $q\in[0.2,0.8]$.
The result is shown in Fig. \ref{two_action_price}.
We can see that $H(x_H)$ is always larger than $H(x_L)$ for a given $q$ which corroborate the monotonicity of the contract result. Furthermore, by increasing the premium fee, the service fee becomes lower. This phenomenon indicates that though the microgrid needs to pay a higher fee to be eligible to receive the resilience service, the service fee is lower once having the membership. The benefits of the microgrid for receiving the resilience service in this scenario are $R(H(x_L),x_L)=\$8.57\rm{K}$ and $R(H(x_H),x_H)=\$2.48\rm{K}$. Thus, the average resilient benefit is $\bar{R}=\$5.525\rm{K}$.

\begin{table}[!t]
\centering
\renewcommand\arraystretch{1.1}
\caption{Value of Parameters in Case Studies\label{t1}}
\begin{tabular}{|c|c|c|c|} \hline
Parameter&Value&Parameter&Value\\ \hline
$P_g^1$& $200\rm{MWh}$& $\tau$ & $60$\\ \hline
$P_g^2$ & $300\rm{MWh}$ & $\alpha$ & $10^{-3}$\\ \hline
$P_g^3$ & $400\rm{MWh}$ & $\beta$ & $30$\\ \hline
$x_1$ & $-150\rm{MWh}$ & $\gamma$ & $100$ \\ \hline
$x_2$& $-100\rm{MWh}$ & $\zeta$& $1500$\\ \hline
$x_3$& $100\rm{MWh}$ & $T$& $\$3\rm{K}$\\ \hline
$x_4$& $140\rm{MWh}$ & $\kappa$& $100$\\ \hline
$x_5$ & $200\rm{MWh}$ & $\rho$ & $1.2$ \\
\hline\end{tabular}
\end{table}

\subsection{Case Study 2: Multiple-Action Multiple-Outcome}
\begin{figure}[!t]
  \centering
  \subfigure[]{
    \includegraphics[width=0.48\columnwidth]{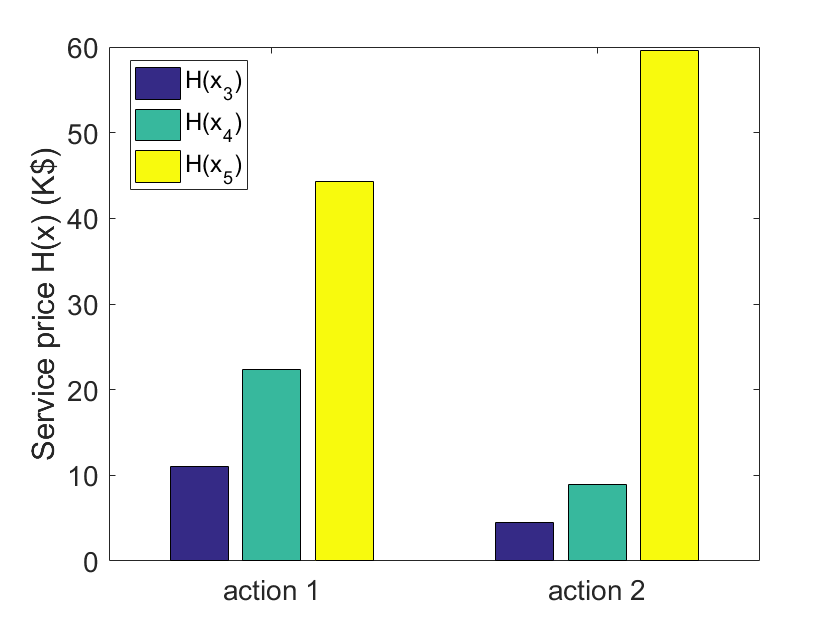}\label{finite_action_2_H}}
	 \subfigure[]{
    \includegraphics[width=0.48\columnwidth]{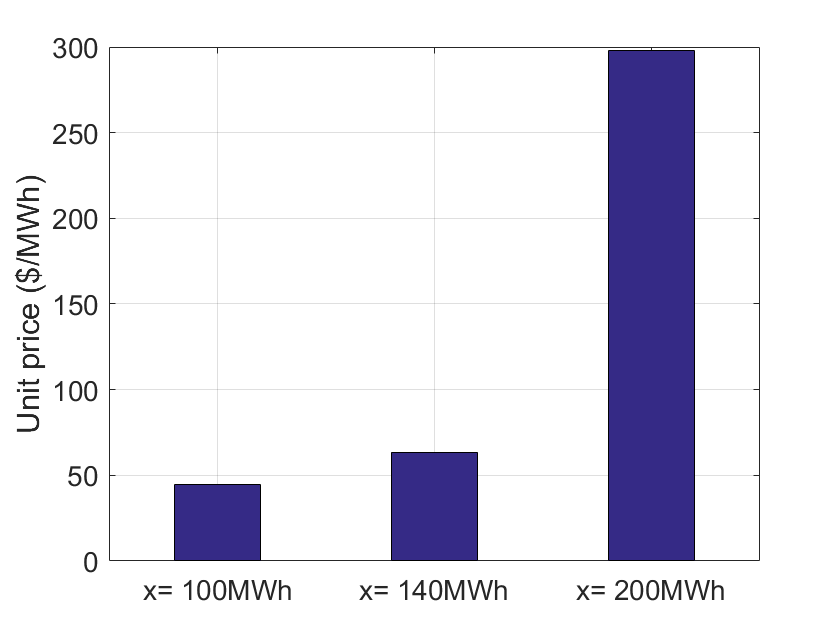}\label{finite_action_2_unit}}
  \caption[]{(a) shows the designed contracts for all implementable actions. (b) shows the optimal resilience contract. With a large premium fee, the resiliency service price in (b) is lower than the one in Fig. \ref{finite_action_2_unit}. }
  \label{finite_action_2}
\end{figure}

We further investigate a more general case study with three actions $P_g^j,\ j\in\mathcal{J}:=\{1,2,3\}$ and five outcomes $x_i,\ i\in\mathcal{N}:=\{1,2,...,5\}$. The future market uncertainty is captured by the distributions below:
\begin{align*}
f(x_i,P_g^1,P_l) = \begin{cases} 
0, & i=1\\
0.1, & i=2\\
0.2, & i=3\\
0.3, & i=4\\
0.4, & i=5
\end{cases},\ &
f(x_i,P_g^2,P_l) = \begin{cases} 
0.1, & i=1\\
0.2, & i=2\\
0.35, & i=3\\
0.35, & i=4\\
0, & i=5
\end{cases},
\end{align*}
\begin{align*}
f(x_i,P_g^3,P_l) &= \begin{cases} 
0.2, & i=1\\
0.3, & i=2\\
0.4, & i=3\\
0.1, & i=4\\
0, & i=5
\end{cases}.
\end{align*}
The premium fee is $T=\$3\rm{K}$, and other parameters are summarized in Table \ref{t1}. 
\begin{figure}[!t]
  \centering
  \subfigure[]{
    \includegraphics[width=0.48\columnwidth]{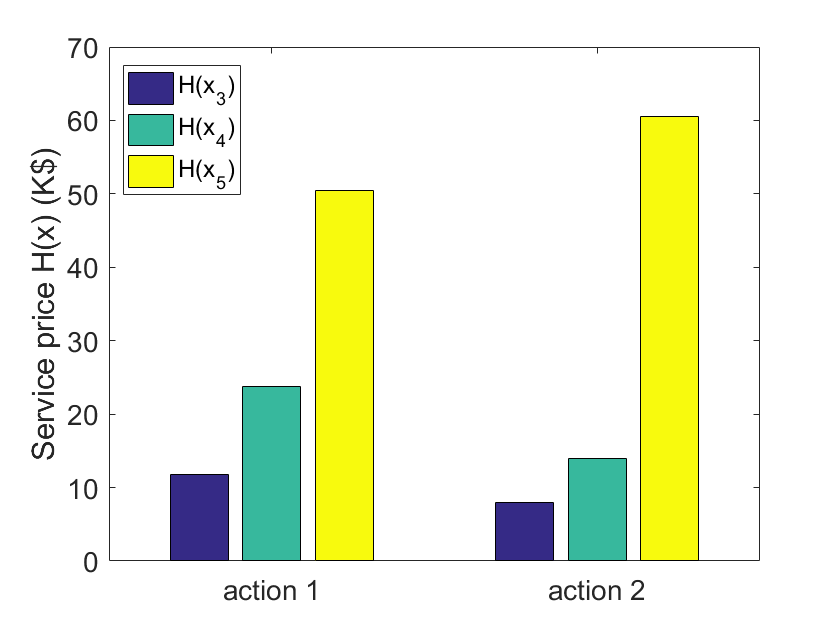}\label{finite_action_1_H}}
	 \subfigure[]{
    \includegraphics[width=0.48\columnwidth]{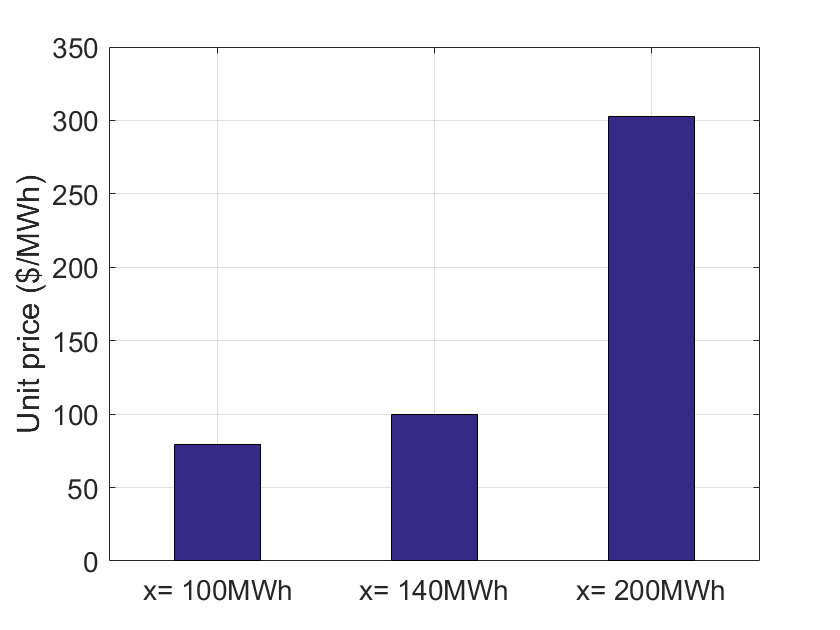}\label{finite_action_1_unit}}
  \caption[]{(a) shows the designed contracts for all implementable actions. (b) shows the optimal resilience contract. In this case, the unit power price increases with the amount of resilience service. }
  \label{finite_action_1}
\end{figure}
We design the resilience contract for this case using the two-step approach presented in Section \ref{finite_action}. First, we need to determine the implementable actions by focusing on the IC and IR constraints. We find that the third action $P_g^3$ will not be adopted by the microgrid under any contract, since the IR constraint does not hold. By designing the optimal contracts for implementing actions $P_g^1$ and $P_g^2$ as shown in Fig. \ref{finite_action_1_H}, we then use the second step in the two-step approach to find the best contract for the SP, and Fig. \ref{finite_action_1_unit} presents the result. In this case, the microgrid (service requester) generates $P_g^2$ amount of power under the optimal contracts provided by the principal. For clarity, we define the unit price of the resilience service as $m(x)=H(x)/x$. Then, the unit power price for the optimal resilience service shown in Fig. \ref{finite_action_1_unit} is $m(x_3)=44.4\rm{\$MW/h}$, $m(x_4)= 63.5\rm{\$/MWh}$ and $m(x_5) = 297.9\rm{\$/MWh}$. In this case, the benefits for receiving the resilience service are $R(H(x_3),x_3)=\$7.24\rm{K}$ and $R(H(x_4),x_4)=\$10.05\rm{K}$. The resulting average resilient benefit is $\bar{R}=\$14.29\rm{K}$. Thus, the proposed RaaS approach has a substantial economic impact in addition to improving the renewable microgrid's resilience.
By increasing the premium fee to $T=\$6\rm{K}$, the results are shown in Fig. \ref{finite_action_2}. In this case, the unit power price is lower than the one in Fig. \ref{finite_action_2_unit}. This is consistent with the observation in the previous case study 1.

\begin{figure}[!t]
  \centering
    \includegraphics[width=0.6\columnwidth]{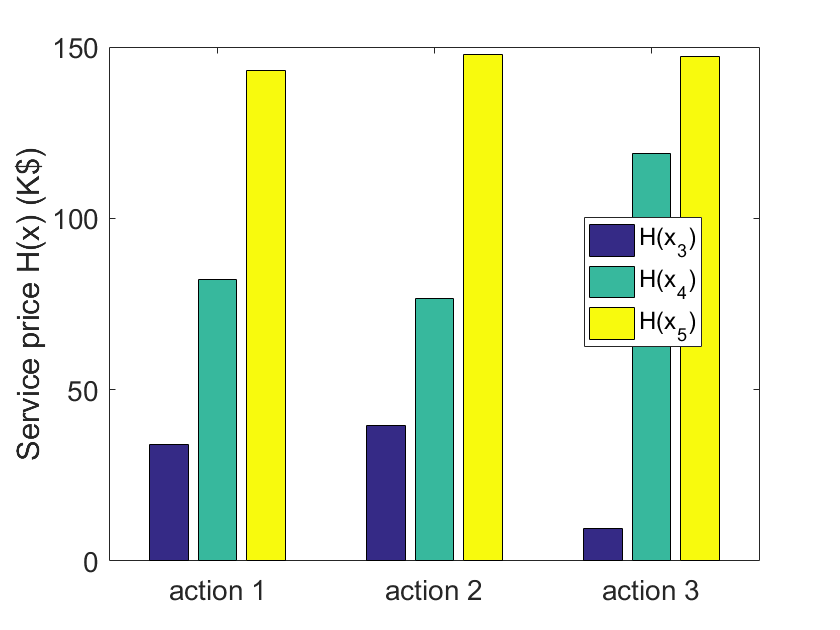}\label{finite_action_3_H}
  \caption[]{The designed contracts for all implementable actions. The optimal contract is the one associated with action 3. The resilience prices for $x_4 = 140\rm{MW}$ and $x_5 = 200\rm{MW}$ are much more expensive than that of  $x_3 = 100\rm{MW}$. }
  \label{finite_action_3}
\end{figure}

When the parameter $\kappa$ in function $\psi$ is changed to $\kappa=250$ which indicates that the agent has a higher payoff by receiving the same amount of resilience service compared with previous case studies, then the contract design results are shown in Fig. \ref{finite_action_3}. Note that all actions are implementable in this scenario. The optimal resiliency contract designed by the SP is the one associated with the implementable action $P_g^3$. We observe that the resilience service prices for $x_4 = 140\rm{MWh}$ and $x_5 = 200\rm{MWh}$ are much more expensive than that of $x_3 = 100\rm{MWh}$. In addition, under the designed optimal contract, the microgrid (service requester) generates $P_g^3$ amount of power, and it is more likely for it to request $x_3 = 100\rm{MWh}$ amount of resilience support from the SP.

\section{Conclusion}\label{conclusion}
In this paper, we have developed a contract-based approach to promote the resilience of interconnected microgrids through a resilience as a service (RaaS) paradigm. Based on principal-agent models that capture the hidden-action nature of the resilience service requester, we have characterized that the service price is monotone when the service requester can take two actions on its local generation. For the finite-action case, we have designed a two-step approach to guide the optimal resiliency contract design. We have also observed that a higher premium of participating this RaaS market decreases the contracted resilience service fee to account for the service requester's incentives. Future work could be extending the current one-principal one-agent model to a networked system with many players and investigating network effects on the RaaS contract design. Another direction of interest is to develop a learning-based approach for the resiliency contract design when the statistics of renewable generation uncertainty is not known.

\bibliographystyle{IEEEtran}
\bibliography{IEEEabrv,references}

\end{document}